\documentclass{PoS}

\title{ITMD factorization and broadening effects in production of forward di-jets}

\ShortTitle{ITMD factorization}

\author{\speaker{Krzysztof Kutak}\\
        Instytut Fizyki Jadrowej Polskiej Akademii Nauk\\ Radzikowskiego 152, 31-342 Krakow\\
        E-mail: \email{krzysztof.kutak@ifj.edu.pl}}

\abstract{I report on the recent result of comparison of forward-forward dijet correlations in azimuthal angle as measured by the ATLAS collaboration in the  proton-proton and proton-lead collisions to calculations within ITMD factorization framework \cite{vanHameren:2019ysa}. The comparison shows that the broadening effect due to interplay of both the gluon saturation and the Sudakov resummation is necessary to describe the data.}

\FullConference{XXVII International Workshop on Deep-Inelastic Scattering and Related Subjects - DIS2019\\
		8-12 April, 2019\\
		Torino, Italy}

\begin{document}

\section{Introduction}

High energy collisions of protons and heavy nuclei at the Large Hadron Collider (LHC) provide a unique tool to probe
dense systems of quarks and gluons.
In particularly interesting are processes where jets or particles are produced 
in the forward direction with respect to the incoming proton. Kinematically, such final  states have large rapidities. 
and therefore they trigger events in which the partons from the nucleus carry rather small longitudinal momentum fraction $x$.
This kinematic setup is perfectly suited to investigate the phenomenon of gluon saturation, which is expected to occur at some value of $x$ to prevent violation of the unitarity bound (for a review of this subject see Ref. \cite{Albacete:2014fwa}). 
In \cite{Aaboud:2019oop} the ATLAS collaboration studied azimuthal correlations of dijets in proton-lead (p-Pb) and proton-proton (p-p) collisions at the center-of-mass energy $\sqrt{s_{NN}}=5.02\,\mathrm{TeV}$ covering the forward rapidity region between $2.7-4.0$ units. The measurement indicates sizable nuclear effects at small values of $x$. 
The behaviour of dense systems of partons when $x$ becomes small is predicted by  Quantum Chromodynamics (QCD) and leads to non-linear evolution equations known as B-JIMWLK equations (for review see \cite{Kovchegov:2012mbw,Gelis:2010nm}), which can be derived within the Color Glass Condensate (CGC) theory.
In CGC, the calculation of forward jet production in dense-dilute collisions relies on the hybrid factorization \cite{Dumitru:2005gt}, where the large-$x$ projectile is described by the collinear PDFs, while the dense target according to theoretical results is described with nonlinear equations. The  description of multi-jet production is rather complicated even in this simplified framework~\cite{Marquet:2007vb}. A novel approach to such processes was initiated in Ref.~\cite{Dominguez:2011wm} for dijets in the back-to-back correlation regime and in Ref.~\cite{Kotko:2015ura} for a more general kinematical configuration. The latter is known as the small-$x$ Improved Transverse Momentum Dependent (ITMD) factorization. The ITMD formula accounts for: 
\begin{itemize}
\item complete kinematics of the scattering process with off-shell gluons, 
\item gauge invariant definitions of the TMD gluon densities, 
\item gauge invariant expressions for the off-shell hard matrix elements, 
\item it also recovers the high energy factorization (aka $k_T$-factorization)~\cite{Catani:1990eg,Collins:1991ty,Deak:2009xt} in the limit of large off-shellness of the initial-state gluon from the nucleus. 
\end{itemize}
Recently, the ITMD  factorization has been proved ~\cite{Altinoluk:2019fui}. Steps in further extension of the formalism to three and more jets were undertaken in Ref.~\cite{Bury:2018kvg} and in \cite{Altinoluk:2018byz} in the correlation limit.
For some of phenomenological application of the formalizm see \cite{vanHameren:2016ftb, Kotko:2017oxg, Albacete:2018ruq}.
While the original ITMD formula, as well as the works studying the jet correlation limit within CGC, include gluon saturation effects, they do not account for all contributions proportional to logarithms of the hard scale set by the  large transverse momenta of jets -- the so-called Sudakov logarithms. It has been shown in Refs.~\cite{vanHameren:2014ala,vanHameren:2015uia} that inclusion of Sudakov logarithms is necessary in order to describe the LHC jet data at small $x$ but yet before the saturation regime. In the low $x$ domain, the resummation leading to the Sudakov logarithms has been developed in ~\cite{Mueller:2012uf,Mueller:2013wwa,Sun:2014gfa,Mueller:2015ael,Zhou:2016tfe,Xiao:2017yya,Zheng:2019zul} see also \cite{Kutak:2014wga}. In the paper \cite{vanHameren:2019ysa}, it has been shown for the first time, that the interplay of saturation effects and the resummation of the Sudakov logarithms is essential to describe the small-$x$ forward-forward di-jet data.
\section{The framework}
The process under consideration is the inclusive dijet production
\begin{equation}
  \mathrm{p} \left(P_{\mathrm{p}}\right) + \mathrm{A} \left(P_{\mathrm{A}}\right) \to j_1 (p_1) + j_2 (p_2)+ X\ ,
\end{equation}
where $A$ can be either the lead nucleus, as in p-Pb scattering, or a proton, as in p-p scattering. 
To describe the above process, we use
the hybrid approach where one assumes that the proton $p$ is a dilute
projectile, whose partons are collinear to the beam and carry momenta $p=x_{\mathrm{p}} P_{\mathrm{p}}$.  
The nucleus $A$ is probed at a dense state. 
The jets $j_1$ and $j_2$ originate from hard partons produced in a collision of the collinear parton $a$
with a gluon belonging to the dense system $A$. This gluon is off-shell, with momentum 
$k=x_{\mathrm{A}} P_{\mathrm{A}} + k_T$ and $k^2=-|\vec{k}_T|^2$.
The ITMD factorization formula for the production of two jets with momenta $p_1$ and $p_2$, and rapidities $y_1$ and $y_2$, reads
\begin{equation}
\frac{d\sigma^{\mathrm{pA}\rightarrow j_1j_2+X}}{d^{2}q_{T}d^{2}k_{T}dy_{1}dy_{2}}
=
\sum_{a,c,d} x_{\mathrm{p}} f_{a/\mathrm{p}}\left(x_{\mathrm{p}},\mu\right) 
\sum_{i=1}^{2}\mathcal{K}_{ag^*\to cd}^{\left(i\right)}\left(q_T,k_T;\mu\right)
\Phi_{ag\rightarrow cd}^{\left(i\right)}\left(x_{\mathrm{A}},k_T,\mu\right)\,,
\label{eq:itmd}
\end{equation}   
The distributions $f_{{a/\mathrm{p}}}$ are the collinear PDFs corresponding to the large-$x$ gluons and quarks in the projectile. 
The functions $\mathcal{K}_{^{ag^*\to cd}}^{_{(i)}}$ are the hard matrix elements constructed from gauge-invariant off-shell amplitudes \cite{vanHameren:2012uj,vanHameren:2012if,Kotko:2014aba,Antonov:2004hh}. 
The quantities $\Phi_{^{ag\rightarrow cd}}^{_{(i)}}$ are the TMD gluon distributions introduced in Ref.~\cite{Kotko:2015ura} and parametrize a dense state of the nucleus or the proton in terms of small-$x$ gluons, see Ref.~\cite{Petreska:2018cbf} for an overview.
The phase space is parametrized in terms of the final state rapidities of jets $y_1,y_2$, as well as the momenta $\vec{k}_T = \vec{p}_{1T}+\vec{p}_{2T}$, and $\vec{q}_T=z\vec{p}_{1T}-(1-z)\vec{p}_{2T}$, where $z=p_1\!\cdot\! P_\mathrm{p}/(p_1\!\cdot\! P_\mathrm{p} + p_2\!\cdot\! P_\mathrm{p})$. The azimuthal angle between the final state partons is
$\Delta\phi$.
The collinear PDFs, hard matrix elements, and the TMD gluon distributions all depend on the factorization/renormalization scale $\mu$. At leading order, the matrix elements depend on $\mu$ only through the strong coupling constant. The collinear PDFs obey the DGLAP evolution when the scale $\mu$ changes. The evolution of the  TMD gluon distributions is more involved. Typically, in saturation physics, one keeps $\mu$ fixed at some scale of the order of the saturation scale $Q_s$, and performs the evolution in $x$ using the B-JIMWLK equation  or its mean field approximation -- the BK equation.
In the present situation, however, we deal with rather hard jets, thus we have $\mu\gg Q_s$ with $Q_s$ being in the perturbative regime $Q_s\gg \Lambda_{\mathrm{QCD}}$. In this kinematic domain, we must account for both $|\vec{k}_T|\sim \mu$ and $|\vec{k}_T|\sim Q_s$ -- the first region corresponds to small $\Delta\phi$, while the second to $\Delta\phi\sim\pi$. In the latter case, the Sudakov logarithms $\ln{(\mu/|\vec{k}_T|)}$ should be resummed. While the perturbative calculation of the Sudakov form factors in the saturation domain has been completed in Ref.~\cite{Mueller:2012uf,Mueller:2013wwa}, in the calculation \cite{vanHameren:2014ala} the Sudakov form factor known from the DGLAP parton showers was used. This procedure corresponds to performing a DGLAP-type evolution from the scale $\mu_0\sim |\vec{k}_T|$ to $\mu$. 
The TMDs entering the formula (\ref{eq:itmd}) for lead and for the proton are constructed from a basic dipole distribution given by the KS gluon density~\cite{Kutak:2012rf} and obtained in~\cite{vanHameren:2016ftb}.
%
\section{Results}
\begin{figure}[t]
  \begin{center}
    \includegraphics[width=0.99\textwidth]{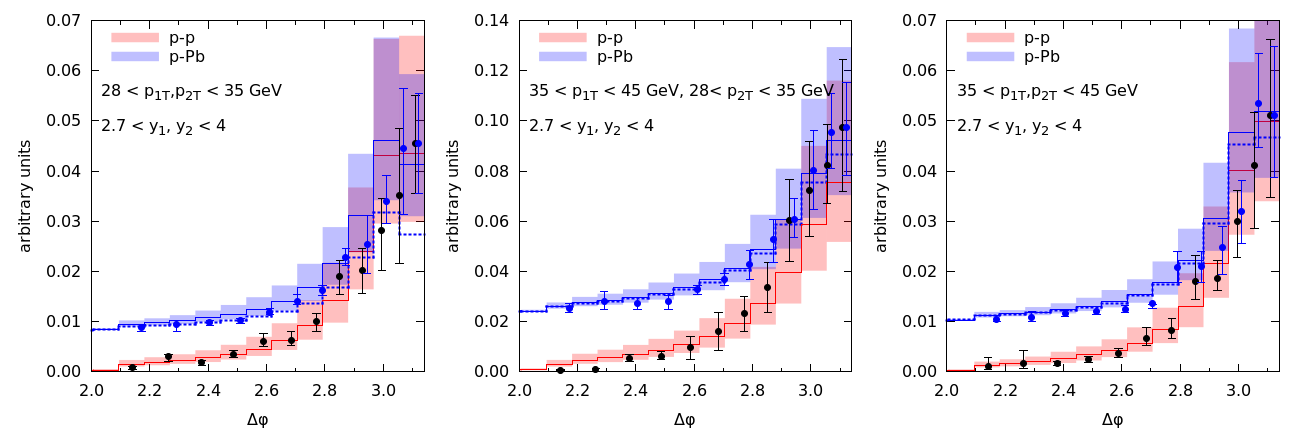}
  \end{center}
  \caption{
    Broadening of  azimuthal decorrelations in p-Pb collisions vs p-p collisions
    for different sets of cuts imposed on the jets' transverse momenta.
    The plots show  normalized cross sections as functions of the azimuthal distance between the two leading jets, $\Delta\phi$. 
    The points show the experimental data \cite{Aaboud:2019oop} for p-p and p-Pb, where the p-Pb data were shifted by a pedestal, so that the values in the bin $\Delta\phi\sim \pi$ are the same.
    Theoretical calculations are represented by the histograms with uncertainty bands coming from varying the scale by factors 1/2 and 2.
    The blue solid and blue dotted lines corresponds to the choices of d=0.5 and d=0.75, respectively.
  }
  \label{fig:broadening}
\end{figure}

Fig.~\ref{fig:broadening}  shows normalized cross sections as functions of $\Delta\phi$ in p-p  and p-Pb collisions. 
The three panels correspond to three different cuts on the transverse momenta of the two leading jets: 
$28<p_{1T}, p_{2T} <35$ GeV,
$35<p_{1T} <45$  and   $28<p_{2T} <35$ GeV, and
$35<p_{1T}, p_{2T} <45$ GeV . 
Both jets are selected in the forward rapidity region, $2.7<y_1,y_2<4.0$, and they are defined with the anti-$k_T$ jet algorithm with the radius $R=0.4$.
The points with error bars represent experimental data from Ref.~\cite{Aaboud:2019oop}. 
It is important to note that the experiment did not measure the cross sections. The experimental points represent the number of two-jet events normalized to the single jet events, as a function of $\Delta\phi$. 
The procedure is as follows. For each set of cuts, a pedestal is added to the p-Pb data, such that the value in the right-most bin (with $\Delta\phi\sim \pi$) is the same as for the p-p data. As seen  in Fig.~\ref{fig:broadening}, the experimental data presented in this way show broadening of the p-Pb distribution.
Theoretical predictions are shown as the red bands for p-p collisions, and the blue bands for p-Pb collisions. The Sudakov resummation described earlier has been included in the predictions.
The same normalization value was then used for the p-Pb predictions.
The main results for p-Pb collisions were obtained with $d=0.5$ and are represented by blue solid lines in Fig.~\ref{fig:broadening}. To estimate the uncertainty associated with the parameter $d$, predictions obtained with the choice $d=0.75$ (in blue lines) are shown.
In order to confront the theoretical broadening effects we obtain from theoretical calculations with those observed experimentally, we add a pedestal to p-Pb results, as determined from the data.
In our framework, the broadening comes from the interplay of the non-linear evolution of the initial state and the Sudakov resummation.
One sees that the theoretical curve describes the shape of the experimental curves well,  within the experimental and theoretical uncertainties, across all the cuts and in the available range of $\Delta\phi$. 
One should emphasize that this is a highly non-trivial consequence of the two components present in our theoretical framework: gluon saturation at low $x$ and Sudakov resummation.
 \section*{Acknowledgement}
The research has been supported by the Polish National Science Centre grant no. DEC-2017/27/B/ST2/01985.
This proceeding is based on results obtained with Piotr Kotko, Andreas van Hameren and Sebastian Sapeta whom I would like to thank for common work.

\bibliographystyle{ieeetr}
\bibliography{references}


\end{document}